\documentclass[fleqn,twoside]{article}
\usepackage{espcrc2}

\newcommand{\AmS}{{\protect\the\textfont2
  A\kern-.1667em\lower.5ex\hbox{M}\kern-.125emS}}

\title{Vacuum Condensates of Dimension Two in Pure Euclidean Yang-Mills}

\author{
D. Dudal\address{Ghent University, Department of Mathematical Physics
 and Astronomy, Krijgslaan 281-S9, B-9000 Gent, Belgium}\thanks{Research
Assistant
 of The Fund For Scientific Research-Flanders,
Belgium}\thanks{david.dudal@rug.ac.be},
A.R. Fazio\thanks{fazio@inp.demokritos.gr}\address{National Research Center
 Demokritos, Ag. Paraskevi, GR-153130 Athens, Hellenic Republic}
V.E.R. Lemes\address{UERJ, Universidade do Estado do Rio de Janeiro, Rua
S{\~a}o
 Francisco Xavier 524, 20550-013 Maracan{\~a}, Rio de Janeiro,
 Brazil}\thanks{vitor@dft.if.uerj.br},
M. Picariello\thanks{marco.picariello@mi.infn.it}\address{Universit{\`a}
degli
 Studi di Milano, via Celoria, 16, I-20133, Milano, and INFN Milano, Italy},
M.S. Sarandy$^{\rm c}$\thanks{sarandy@dft.if.uerj.br},
S.P. Sorella$^{\rm c}$\thanks{sorella@uerj.br}\thanks{Talk given by S.P.
Sorella
 at the International Conference {\it Renormalization Group and Anomalies in
 Gravity and Cosmology}, 17-23 March, 2003, Ouro Preto, MG, Brazil}, and
H. Verschelde$^{\rm a}$\thanks{henri.verschelde@rug.ac.be}}

\begin{document}

\begin{abstract}

Gluon and ghost condensates of dimension two and their relevance for
Yang-Mills theories are briefly reviewed.
\vspace{1pc}
\end{abstract}

\maketitle

\section{The gauge condensate {\ $\left\langle A^{2}\right\rangle $ in the
Landau gauge}}

\subsection{Motivation}

{We shall consider pure Euclidean $SU(N)$ Yang-Mills }
\[
S_{YM}=-\frac{1}{4}\int d^{4}xF_{\mu \nu }^{a}F_{\mu \nu }^{a}
\]
\noindent {\ In the last few years lattice simulations \cite{lat}
of the two and three point functions of $SU(N)$ Yang-Mills in the
Landau gauge have reported the existence of a large discrepancy
between the expected perturbative behavior and the lattice
results. The discrepancy is sizeable up to energies $\approx $
}${10GeV}${, which is a rather big value compared to $\Lambda
_{QCD}\approx $}${(200-300)MeV}${. According to 
\cite{lat,gz}, the discrepancy could be explained by adding to the
perturbative result a power correction of the kind $1/k^{2},$ by
introducing the dimension two gauge condensate $\left\langle
A^{2}\right\rangle =$ $\left\langle A_{\mu }^{a}A_{\mu
}^{a}\right\rangle $}, namely {\
\[
k^{2}G^{(2)}(k^{2})=G_{{\rm PERT}}^{(2)}(k^{2})+c\frac{\left\langle
A^{2}\right\rangle }{k^{2}}
\]
\[
G^{(2)}=\frac{\delta _{ab}}{3(N^{2}-1)}\left( \delta _{\mu \nu }-\frac{%
k_{\mu }k_{\nu }}{k^{2}}\right) {\left\langle A_{\mu }^{a}(k)A_{\nu
}^{b}(-k)\right\rangle }
\]
\noindent and
\[
\alpha _{{\rm run}}(k^{2})=\alpha _{{\rm PERT}}(k^{2})+c^{\prime }\frac{%
\left\langle A^{2}\right\rangle }{k^{2}}
\]
where the coefficients $c$, $c^{\prime }$ are obtained through OPE \cite{ope}. The
lattice estimate for the gauge condensate is }$\left\langle
A^{2}\right\rangle \approx \left( 1.64GeV\right) ^{2}\;$at the energy scale$%
{\rm \;}\mu =10GeV$ {\cite{lat}. The existence of a nonvanishing condensate $%
\left\langle A^{2}\right\rangle $ could be deeply related to the dynamical mass
generation for the gluons and to the instability of the causal perturbative Yang-Mills
vacuum \cite{savv}. Lattice results \cite{mg1} have indeed
reported something like }$m_{{\rm gluon}}\approx 600MeV$.
It is worth mentioning that theoretical analysis of the gluon propagator in the Landau gauge have
shown that its behavior is suppressed in the infrared region \cite{gr,zw,sd,ksupp,bl,zw2}, in agreement
with lattice simulations \cite{mg1,mg2,cucch}.
{\ The gauge condensate $\left\langle A^{2}\right\rangle $ might also be relevant for
confinement \cite{ck}, as it could lead to the area law for the vacuum expectation value of
the Wilson loop }$W\sim \exp \left( -\sigma {\rm Area}\right) $ with $\sigma
\sim ${$\left\langle A^{2}\right\rangle $.}

\noindent In particular, as underlined in \cite{gz}, $\left\langle
A^{2}\right\rangle $ should receive contributions from both long and short
distances, {\it i.e.}
\[
\left\langle A^{2}\right\rangle =\left\langle A^{2}\right\rangle _{{\rm LD}%
}+\left\langle A^{2}\right\rangle _{{\rm SD}}\;.  \label{ldsd}
\]
For what concerns the long distance part $\left\langle A^{2}\right\rangle _{%
{\rm LD}}$,  Ph. Boucaud et al. {\cite{ist}} have established that instantons do
contribute to $\left\langle A^{2}\right\rangle _{%
{\rm LD}}$. The lattice estimate of the instanton contribution has been found
$\left\langle A^{2}\right\rangle _{{\rm INST}%
}\approx (1.7)GeV^{2}$. For the short distance contribution $\left\langle
A^{2}\right\rangle _{{\rm SD}}$, H. Verschelde et al. \cite{hv} have been
able to obtain the two-loop effective potential for $\left\langle
A^{2}\right\rangle $ by combining the Local Composite Operators technique
with the Renormalization Group Equations. They obtained a gap equation whose
weak coupling solution yields a nonvanishing condensate, resulting in a
gluon mass $m_{{\rm gluon}}\approx 500MeV$.

\subsection{Why $A^{2}$ in the Landau gauge}

A simple naive argument shows that $\int d^{4}xA^{2}$ is invariant under
infinitesimal gauge transformations in the Landau gauge $\partial A=0$,
namely
\[
\delta A_{\mu }^{a}=-\left( D_{\mu }\omega \right) ^{a}
\]
\[
\delta \int d^{4}x\frac{1}{2}A^{2}=\int d^{4}x\omega ^{a}\partial A^{a}=0
\]
In the BRST framework we have that, in the Landau gauge, $\int d^{4}xA^{2}$
is BRST invariant on shell
\[
s\int d^{4}xA^{2}=0{\ }+{\ }{\rm eqs.{\ }{\ }of{\ }{\ }motion}
\]
A more precise meaning for $A^{2}$ is provided by introducing
the nonlocal gauge invariant operator $A_{{\rm \min }}^{2}$, obtained by minimizing  $%
\int d^{4}xA^{2}$ along the gauge orbits, namely
\[
A_{{\rm \min }}^{2}{\bf =}{\rm \;}\left[ {\rm \min_{\{U\}}.\;}\int
d^{4}x\,\left( A_{\mu}^{U} \right)^2 \; \right]
\]
where $U$ denotes a generic gauge transformation. Of course, {$A_{{\rm \min }}^{2}$ is
stationary under gauge transformations. Furthermore, the
minimum condition for $\int d^{4}x\,A^{2}$ is given by the Landau gauge $%
\partial A=0$. A deep relationship between $\int d^{4}x\,A^{2}$ and $A_{{\rm %
\min }}^{2}$ is thus expected to hold in the Landau gauge. In fact, as discussed in
\cite{gsz}, it turns out that in the abelian case $\int d^{4}xA^{2}=A_{{\rm %
\min }}^{2}$. Also, from \cite{gsz}, one learns that the condensate $%
\left\langle A^{2}\right\rangle $ can be used as a useful order parameter for the phase
transition of compact $QED$ in $3D$. }In the nonabelian case, {the
situation is more complex. It is true that the Landau gauge condition $%
\partial A=0$ is a stationary condition for the functional $\int
d^{4}x\,A^{2}$. However, in this case, one has to face the existence of
Gribov's ambiguities for large values of the gauge field \cite{zw}. A recent discussion about $A_{{\rm \min }}^{2}$ and Gribov's ambiguities can be found in \cite{vb}. }

\noindent The operator {\ $A^{2}$ displays also remarkable ultraviolet properties.
It is  multiplicatively renormalizable, its anomalous dimension $%
\gamma _{\,A^{2}}$ being available up to three loops in the
$\overline{MS}$  scheme \cite{jg}. Recently, it has
been proven \cite{dvs} by using BRST Ward identities that $\gamma
_{\,A^{2}}$ is not an independent parameter of the theory,

being expressed as a combination of the gauge beta function $\beta $ and of
the anomalous dimension $\gamma _{A}$ of the gauge field $A$, according to
the relationship
\[
\gamma _{\,A^{2}}=-\left( \frac{\beta (a)}{a}+\gamma _{A}\right)
\;,\;\;\;\;a=\frac{g^{2}}{16\pi ^{2}}\;.
\]
} {\ }

\section{{{{Generalization of \ $\left\langle A^{2}\right\rangle $ to {\bf %
other gauges }}}}}

\subsection{\bf The Maximal Abelian Gauge }

The Maximal Abelian gauge (MAG) plays an important role for the dual superconductivity picture
for confinement based on the electromagnetic duality proposed by \cite{ed}.
This gauge is extensively used in lattice simulations.
It has provided evidences \cite{sy} for the Abelian dominance hypothesis \cite{adom} and for
monopoles condensation \cite{mc}. In the MAG, the gauge field is decomposed according
to the generators of the Cartan subgroup of the gauge group. For $SU(2)$%
\[
A_{\mu }^{a}T^{a}=A_{\mu }T^{3}+A_{\mu }^{\alpha }T^{\alpha
}\;,\;\;\;\;\;\alpha =1,2
\]
For the gauge fixing we have
\[
\int d^{4}x\left[ \frac{1}{2\xi }F^{\alpha }F^{\alpha }-\overline{c}^{\alpha
}M^{\alpha \beta }c^{\beta }-g^{2}\xi \left( \overline{c}^{\alpha
}\varepsilon ^{\alpha \beta }c^{\beta }\right) ^{2}\right]
\]
where $\xi$ denotes the gauge parameter and
\[
F^{\alpha }=D_{\mu }^{\alpha \beta }A_{\mu }^{\beta }=\left( \partial _{\mu
}A_{\mu }^{\alpha }+g\varepsilon ^{\alpha \beta }A_{\mu }A_{\mu }^{\beta
}\right)
\]
with
\[
M^{\alpha \beta }=D_{\mu }^{\alpha \gamma }D_{\mu }^{\gamma \beta
}+g^{2}\varepsilon ^{\alpha \gamma }\varepsilon ^{\beta \sigma }A_{\mu
}^{\gamma }A_{\mu }^{\sigma }
\]
The MAG allows for a residual local $U(1)$ invariance, which has
to be fixed later on. It is a nonlinear gauge. As a consequence, a
quartic ghost interaction has to be introduced for consistency
\cite{cons}. Lattice simulations have shown that the off-diagonal
components $A_{\mu }^{\beta }$ acquire a mass \cite{as,b}. These
components should decouple at low energies, according to the
Abelian dominance. Therefore, the understanding of the mechanism
for the dynamical mass generation for the off-diagonal components
is fundamental for the Abelian dominance. It is remarkable that
the operator $A^{2}$ of the Landau gauge can be generalized to the
MAG \cite{kmsi2,kmsi}. The gluon-ghost dimension two
operator
\[
O_{{\rm MAG}}=\left( \frac{1}{2}A_{\mu }^{\alpha }A_{\mu }^{\alpha }+\xi
\overline{c}^{\alpha }c^{\alpha }\right)
\]
has indeed the following property
\[
s\int d^{4}x\;O_{{\rm MAG}}=0\;+\;{\rm eqs.\;of \; motion}
\]
The condensate $\left\langle O_{{\rm MAG}}\right\rangle $ should
play a very important role for the Abelian dominance, as it would provide
effective masses for the off-diagonal components. However, at present, very
little is known about the operator $O_{{\rm MAG}}$ and the possible
existence of $\left\langle O_{{\rm MAG}}\right\rangle $. Concerning the UV
properties of $O_{{\rm MAG}}$, it has been proven to be multiplicatively
renormalizable \cite{ew}.

\subsection{\protect\vspace{3mm}{{\bf The Curci-Ferrari gauge } }}

{The so called Curci-Ferrari gauge resembles very much the MAG. It can
thus provide useful insights about the gluon-ghost condensate. For the
gauge fixing we have now
\begin{eqnarray*}
&& \hspace{-.8cm}\int d^{4}x \left( \frac{1}{2\xi }\left( \partial A^{a}\right) ^{2}+%
\overline{c}^{a}\partial _{\mu }D_{\mu }c^{a}+\frac{\xi g}{2}f^{abc}\partial
A^{a}\overline{c}^{b}c^{c}   \right. \\
&& \left. -\frac{\xi g^{2}}{16}f^{abc}\overline{c}^{a} \overline{c}^{b}f^{mnc}c^{m}c^{n}\right) \;
\end{eqnarray*}
where $a=1,...., (N^{2}-1),$ for $SU(N)$. Notice the presence of
the quartic ghost term. The operator $O_{{\rm MAG}}$ generalizes
 \cite{kmsi2,kmsi} to the CF gauge as
\[
O_{{\rm CF}}=\left( \frac{1}{2}A_{\mu }^{a}A_{\mu }^{a}+\xi \overline{c}%
^{a}c^{a}\right)
\]
and
\[
s\int d^{4}x\;O_{{\rm CF}}=0\;+\;{\rm eqs.\; of \; motion}
\]
Some properties of the operator $O_{{\rm CF}}$ are known. $O_{{\rm
CF}}$ is  multiplicatively renormalizable \cite{kmsi}.
Its anomalous dimension has been computed till three loops in the
$\overline{MS}$ scheme \cite{jg}. Recently, the
effective potential for $O_{{\rm CF}}$ has been obtained in
\cite{dvlssp}, yielding a gap equation whose weak coupling
solution gives a nonvanishing condensate $\left\langle O_{{\rm
CF}}\right\rangle $, resulting in a dynamical mass generation.
This gives an indication that something similar should happen in
the MAG. It is also worth remarking that the Landau gauge, the MAG
and the CF gauge have many features in common. All these gauges
possess a larger set of global symmetries, giving rise to the so
called Nakanishi-Ojima (NO)\ algebra \cite {dno}. The operators
$A^{2},\;$ $O_{{\rm MAG}},\;O_{{\rm CF}}$ are left
invariant\footnote{Invariant on-shell for what
concerns the (anti-)BRST.} by the NO algebra. }

\section{\bf Evidences for ghost condensates }

{Contrary to the gauge condensate $\left\langle A^{2}\right\rangle $, the
first proposal for the ghost condensation has been made in the Maximal
Abelian gauge by \cite{ms,kk}. Due to the
quartic ghost-antighost self interaction of the MAG
\[
\left( \overline{c}^{\alpha }\varepsilon ^{\alpha \beta }c^{\beta }\right)
^{2}
\]
ghosts might condense, giving rise to bound states. To some extent, the
mechanism is similar to the formation of fermion bound states in the Nambu
Jona-Lasinio model. Several channels for the ghost condensates are possible
\cite{ch}, corresponding to different values of the Faddeev-Popov charge, namely
\[
\left\langle \overline{c}^{\alpha }\varepsilon ^{\alpha \beta }c^{\beta
}\right\rangle \;\;\;\;\;\;{{\rm Faddeev-Popov\ \ charge\ 0}}
\]
\[
\left\langle c^{\alpha }\varepsilon ^{\alpha \beta }c^{\beta }\right\rangle
\;\;\;\;\;\;{\rm Faddeev-Popov \;charge\;+2\;}
\]
\[
\left\langle \overline{c}^{\alpha }\varepsilon ^{\alpha \beta }\overline{c}%
^{\beta }\right\rangle \;\;\;\;\;\;{\rm Faddeev-Popov \;charge\;-2}
\]
\noindent The existence of several channels for the ghost condensation
can be related to the dynamical symmetry breaking of the generators of the $%
SL(2,R)$ subalgebra of the NO algebra \cite{dno}. It has an interesting
analogy with ordinary superconductivity, known as the BCS versus Overhauser
effect. The BCS channel corresponds to the charged particle-particle and hole-hole pairing,
while the Overhauser to the particle-hole pairing. In the present case the
Faddeev-Popov charged condensates $\left\langle c^{\alpha }\varepsilon ^{\alpha \beta
}c^{\beta }\right\rangle $, $\;\left\langle \overline{c}^{\alpha
}\varepsilon ^{\alpha \beta }\overline{c}^{\beta }\right\rangle $ would
correspond to the BCS channel, while $\left\langle \overline{c}^{\alpha
}\varepsilon ^{\alpha \beta }c^{\beta }\right\rangle $ to the Overhauser
channel.

Evidences for the existence of the ghost condensates have been reported
also in the Curci-Ferrari gauge \cite{kcf,lcf,hs}. Although the
quartic ghost-antighost interaction is absent in the Landau gauge, it has
been possible by combining the Local Composite Operators technique with the
Algebraic Renormalization to give evidences for the ghost condensation in
this gauge \cite{ll}.

\noindent Many aspects of the gauge and ghost condensation are under investigation \cite{app}, deserving a deeper understanding.
Some of them are:

$\bullet$ Analysis of the BCS versus Overhauser effect and its relationship
with the breaking of the NO algebra, present in MAG, CF and Landau gauge.

$\bullet$ The role of the color and BRST symmetry in the presence of the gauge
and ghost condensates.

$\bullet$ Modification of the infrared behavior of the ghost propagator and possible
consequences for the Schwinger-Dyson equations. The ghost condensation modifies indeed the
off-diagonal ghost propagator in the infrared as
\[
\left\langle \overline{c}^{\alpha }(k)c^{\beta }(-k)\right\rangle =\frac{%
k^{2}\delta ^{\alpha \beta }+v\varepsilon ^{\alpha \beta }}{k^{4}+v^{2}}
\]
while for the diagonal component
\[
\left\langle \overline{c}^{3}(k)c^{3}(-k)\right\rangle =\frac{1}{k^2}
\]
where $v$ stands for the value of the condensation.
As underlined in \cite{dvmg}, both gauge and ghost condensates {$\left\langle A^{2}\right\rangle $, }$%
\left\langle \overline{c}^{\alpha }\varepsilon ^{\alpha \beta }c^{\beta
}\right\rangle $, $\left\langle c^{\alpha }\varepsilon ^{\alpha \beta
}c^{\beta }\right\rangle $, $\left\langle \overline{c}^{\alpha }\varepsilon
^{\alpha \beta }\overline{c}^{\beta }\right\rangle $ might play an
important role for a better understanding of the nature of the mass gap in
Yang-Mills theories.

\section*{Acknowledgments}

S.P. Sorella is grateful to the Organizing Committee
of the Conference for the kind invitation and to M. Asorey for
many valuable and helpful discussions. The Conselho Nacional de
Desenvolvimento Cient\'{\i }fico e Tecnol\'{o}gico CNPq-Brazil,
the Funda{\c{c}}{\~{a}}o de Amparo {a Pesquisa do Estado do Rio de
Janeiro (Faperj), the SR2-UERJ and the MIUR-Italy are acknowledged
for the financial support. }The work of A.R. Fazio was supported
by EEC Grant no. HPRN-CT-1999-00161 and partially by Fondazione
''Angelo Della Riccia''-Ente morale.

\end{document}